# Giant topological Hall effect in correlated oxide thin films

Lorenzo Vistoli[1], Wenbo Wang[2], Anke Sander[1♦], Qiuxiang Zhu[1♦], Blai Casals[3], Rafael Cichelero[3], Agnès Barthélémy[1], Stéphane Fusil[1], Gervasi Herranz[3], Sergio Valencia[6], Radu Abrudan[6], Eugen Weschke[6], Kazuki Nakazawa[4], Hiroshi Kohno[4], Jacobo Santamaria[5], Weida Wu[2], Vincent Garcia[1] and Manuel Bibes[1]*

[1] Unité Mixte de Physique, CNRS, Thales, Université Paris-Saclay, 91767 Palaiseau, FRANCE

[2] Department of Physics and Astronomy, Rutgers University, Piscataway, New Jersey 08854, USA

[3] Institut de Ciència de Materials de Barcelona (ICMAB-CSIC), Campus de la UAB, 08193 Bellaterra, Catalonia, SPAIN

[4] Department of Physics, Nagoya University, Nagoya 464-8602, JAPAN

[5] GFMC, Dpto. Física de Materiales, Universidad Complutense de Madrid, 28040, SPAIN

[6] Helmholtz-Zentrum Berlin für Materialen & Energie, Albert-Einstein-Strasse 15, 12489 Berlin, GERMANY

Strong electronic correlations can produce remarkable phenomena such as metal-insulator transitions[1] and greatly enhance superconductivity[2], thermoelectricity[3], or optical non-linearity[4]. In correlated systems, spatially varying *charge* textures also amplify magnetoelectric effects[5] or electroresistance in mesostructures[6]. However, how spatially varying *spin* textures may influence electron transport in the presence of correlations remains unclear. Here we demonstrate a very large topological Hall effect (THE)[7,8] in thin films of a lightly electron-doped charge-transfer insulator, (Ca, Ce)MnO₃. Magnetic force microscopy reveals the presence of magnetic bubbles, whose density *vs.* magnetic field peaks near the THE maximum, as is expected to occur in skyrmion systems[9]. The THE critically depends on carrier concentration and diverges at low doping, near the metal-insulator transition. We discuss the strong amplification of the THE by correlation effects and give perspectives for its non-volatile control by electric fields.

\* manuel.bibes@cnrs-thales.fr

♦ these authors contributed equally to the manuscript.

The role of topology in condensed matter is the subject of a strong research effort that has led to the discovery of distinct electronic phases and novel physical phenomena in the last decade[10]. In *reciprocal* space, strong spin-orbit coupling may produce topological insulators with exotic band structures[11]. Electronic correlations strongly enrich the diversity of possible topological phases in such quantum materials[10], and various novel states including topological Mott insulators, axion insulators or quantum spin liquids have been predicted, mainly in transition metal oxides[12]. In *real* space, topological defects in a spin lattice such as skyrmions (particle-like spin configurations characterised by an integer winding number or topological charge $Q \neq 0$)[13,14] give rise to emerging phenomena such as the topological Hall effect (THE)[7,8]. However, while several oxides also exhibit topological spin textures[14–16], how these couple to the strong correlations (ubiquitous in these materials) to influence electron transport has not been addressed up to now.

In this paper, we explore the interplay between strong correlations and micromagnetic configurations in perpendicularly magnetized epitaxial thin films of a lightly doped charge-transfer insulator[17], CaMnO₃. Undoped CaMnO₃ is a Pbnm perovskite in which magnetic order is dominated by first-neighbour antiferromagnetic super-exchange interactions, with an additional small ferromagnetic component along z ($G_xA_yF_z$ order[18] ; x, y and z are parallel to the axes of the orthorhombic unit cell). A sketch of this spin structure is shown in Fig. 1a. This small moment is about 0.04 $\mu_B$/Mn, corresponding to a canting angle of about 1° and the Néel temperature of bulk CaMnO₃ is 120 K.

Remarkably, upon doping by just a few percent of Ce⁴⁺ at the Ca site, the material transitions to a metallic state[19] and the weak moment increases up to $M_S \approx 0.7$ $\mu_B$/Mn, signalling the onset of ferromagnetic double exchange associated with carrier delocalisation. In thin films grown on YAlO₃, the transition from an insulating to a metallic state occurs at Ce concentrations as low as ~1-2% (Ref.[20]). In addition, on this substrate imposing a compressive strain of ~1%, magnetoelastic anisotropy induces an easy magnetization axis perpendicular to the film plane[20]. In such a perpendicular magnet where the condition $\frac{2K_u}{\mu_0 M_S^2} \gg 1$ is easily met ($K_u$ is the magnetic anisotropy and $M_S$ the magnetization at high field), magnetization should reverse through the formation of magnetic rod-like domains often referred to as bubbles. Interestingly, such bubble-like domains may exhibit topological properties[7] akin to those of skyrmions albeit without the need for bulk or interfacial inversion symmetry breaking usually required for skyrmion formation. In fact, topological bubbles[7] with diameters in the 100-300 nm range have been observed in several bulk *centrosymmetric* manganites[14–16]. The unique coexistence of metallicity and perpendicular magnetic anisotropy in a doped charge-transfer insulator thus makes (Ca, Ce)MnO₃ (CCMO) a well-suited system to probe how topological magnetic configurations may influence electron transport in the presence of correlations.

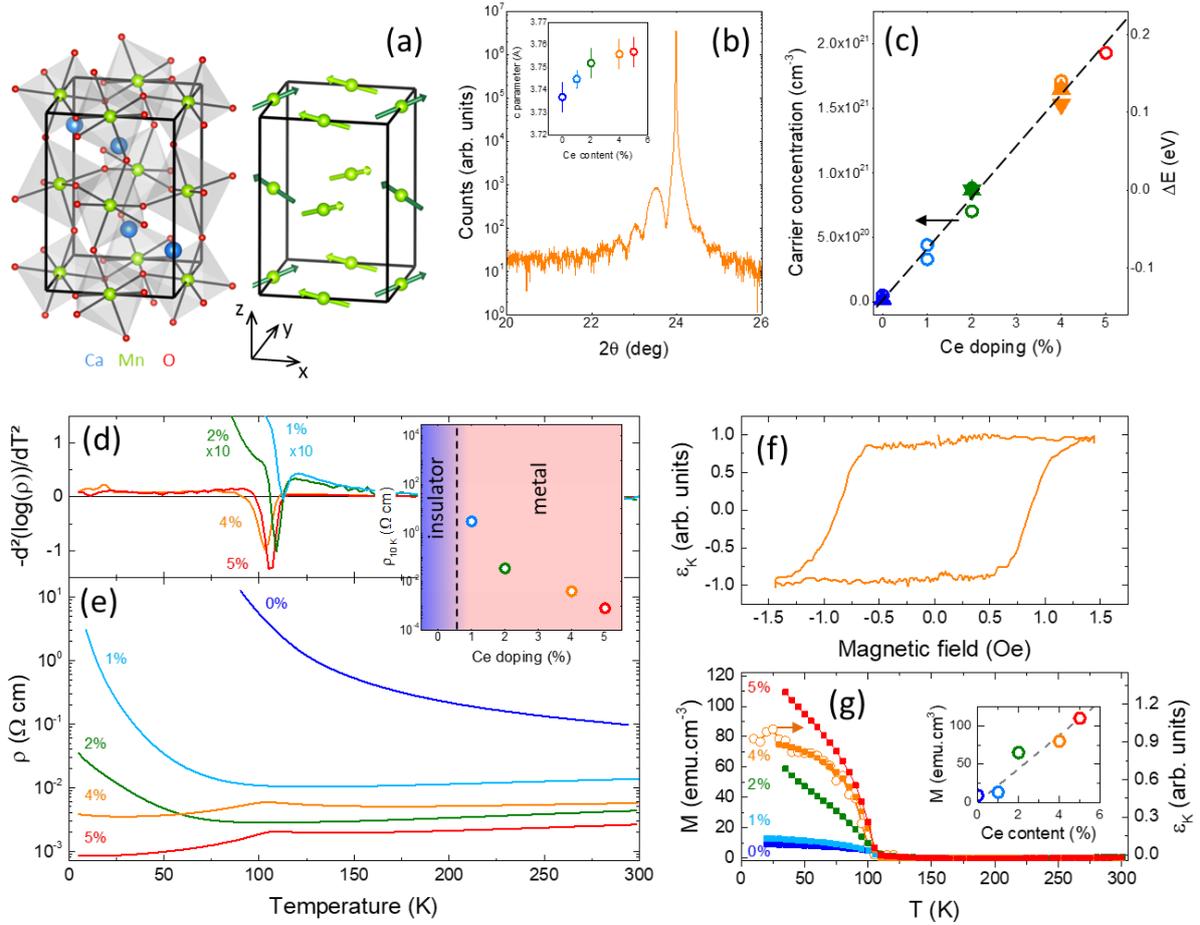

**Figure 1. Doping-dependant structural, electronic and magnetic properties of Ca$_{1-x}$Ce$_x$MnO$_3$ thin films.** (a) Sketch of the structural and spin structure of CCMO in the orthorhombic unit cell. In the right panels, the light green arrows point towards -y, and the dark green ones towards +y. The canting angles are exaggerated for clarity. (b) 2θ-ω scan near the pseudocubic (001) reflection of a x=4% CCMO film on YAO(001). Inset : doping dependence of the out-of-plane lattice parameter c. (c) Doping dependence of the carrier density (open circles, left axis) and of the energy shift of the Mn L$_3$ peak (relative to the position at 2% doping) from EELS (up triangles, reproduced from Ref. [21]) and XAS (down triangle). Temperature dependence of the resistivity (e) and the second derivative of its logarithm (d) for different doping levels. The phase diagram in the inset shows the dependence of the resistivity at low temperature. (f) Kerr ellipticity vs magnetic field at 15 K for a 4% film. (g) Temperature dependence of the magnetization in a field of 1 kOe (left axis, solid symbols) for different doping levels and of the Kerr ellipticity (right axis, open symbols) for a 4% film. The inset shows the doping dependence of the magnetization.

CCMO epitaxial thin films with nominal Ce concentration x=0, 1%, 2%, 4% and 5% were grown by pulsed laser deposition on (001)-oriented YAlO$_3$ single crystal substrates. Fig. 1b presents a typical X-ray diffraction $2\theta$-$\omega$ scan for a 4% sample, with Laue fringes attesting of the structural coherence of the film. The inset shows that the out-of-plane parameter (collected on fully strained films) increases with doping, as expected in a solid solution. The Ce doping modifies the Mn valence from $Mn^{4+}$ in pure $CaMnO_3$ and introduces $Mn^{3+}$, as revealed by a shift towards low energies of EELS (electron energy loss spectroscopy) and XAS (X-ray absorption spectroscopy) spectra measured at the Mn $L_3$ edge (see Fig. 1c and Ref. [21]). Accordingly, the carrier concentration also varies linearly with Ce doping, (Fig. 1c, left axis).

Fig. 1e shows the temperature ($T$) dependence of the longitudinal resistivity $\rho$ for different doping levels. While the pure $CaMnO_3$ sample exhibits a high resistivity and a thermally activated behaviour − in line with its expected charge-transfer insulating state − Ce doping yields a decrease of the resistivity by several orders of magnitude and a room-temperature metallic response already for 1% Ce (CCMO1). At 4% (CCMO4) and 5% (CCMO5) the resistivity shows a cusp at ∼110K that signals the transition to the weak-ferromagnetic state. A signature of this transition is also present in the CCMO1 and CCMO2 data and translates into a dip in the temperature dependence of $d^2\log(\rho)/dT^2$ (cf. Fig. 1d), indicating that these samples also possess a weak-ferromagnetic behaviour. The inset presents the doping dependence of the low temperature resistivity (open circles).

The weak-ferromagnetic character of the films is evident from the temperature dependence of the magnetization $M$ (and of the Kerr ellipticity for x=4%) displayed in Fig. 1g. For all doping levels, the $T_C$ is close to 110 K but $M$ increases with doping, as in the bulk[22] (see inset). Note that these data were measured with a modest out-of-plane field of 0.1 T, so that the shape of the $M$ vs $T$ curves is consistent with a perpendicular magnetic anisotropy. This is also illustrated in Fig. 1f that shows the Kerr ellipticity vs out of plane magnetic field for a CCMO4 sample; a remanence of virtually 100% is observed. Finally, the centrosymmetric nature of the films was confirmed by the absence of a signal in second harmonic generation.

To probe the influence of the micromagnetic structure on the transport response, we have measured the Hall effect as a function of temperature. The results for CCMO4 are summarized in Fig. 2. Fig. 2a presents the transverse (Hall) resistivity at different temperatures. At 130 K the Hall effect is linear, consistent with n-type transport. The carrier density is $n \approx 1.7\ 10^{21}$ cm$^{-3}$, close to the nominal value of $1.55\ 10^{21}$ cm$^{-3}$ for 4% Ce, assuming that each Ce ion brings two electrons. As temperature decreases below $T_C$ (see the 95 K data), a hysteretic component develops, corresponding to the anomalous Hall effect (AHE) in a perpendicularly magnetised sample. As temperature decreases further, a third component appears in the form of a peak centred in the 1 T range. Its peculiar shape is reminiscent of the topological Hall effect (THE) observed in skyrmion systems[8,23–25]. In these samples, the Hall effect thus comprises three components, i.e.

$$\rho_{Hall} = R_0 H + R_S M + \rho_{THE} \qquad (1)$$

To obtain the AHE and the THE from the Hall data, we first subtract a high field slope $R_0$ (ordinary Hall effect). To extract the AHE (second term in Eq.(1), i.e. $\rho_{AHE} = R_S M$) we use magneto-optical Kerr effect measurements on the same sample through the field dependence of the Kerr ellipticity (that is expected to be proportional to the magnetization). This decomposition is shown in Fig. 2b. The amplitude of the THE is comparable to that of the AHE and much higher than previously reported values in other materials already for this 4% sample, cf. Table 1. Fig. 2c shows the extracted THE component for different temperatures, confirming that the THE is in the μΩ.cm range at low temperature and decreases upon heating to vanish between 75 and 95 K.

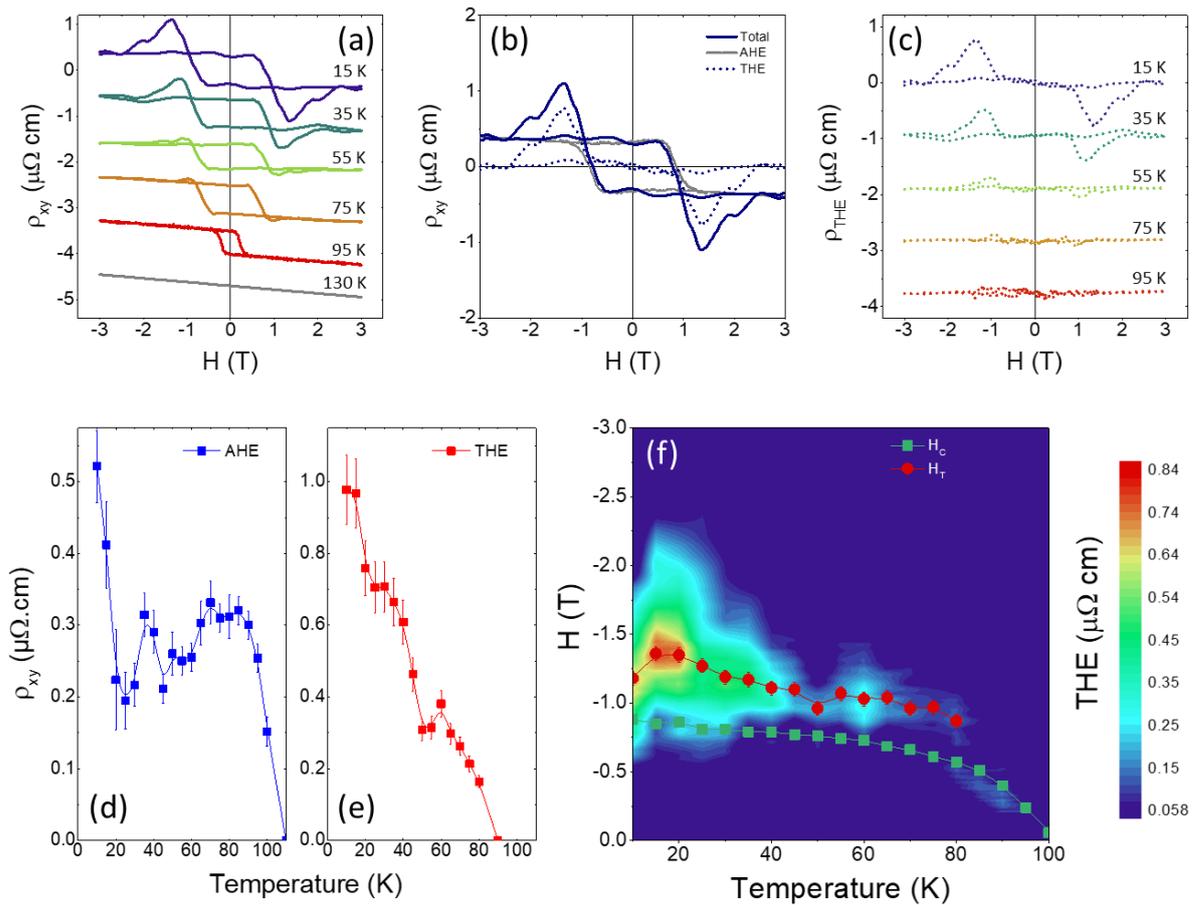

**Figure 2. Topological Hall effect in 4% CCMO.** (a) Hall effect at different temperatures. The data are shifted vertically for clarity. (b) Decomposition of the Hall effect into the AHE and the THE using Kerr ellipticity data. (c) Topological Hall effect at different temperatures. As in (a), the data are shifted vertically for clarity. Temperature dependence of the anomalous Hall resistivity (d) and the topological

*Hall resistivity (e). (f) Topological Hall effect vs temperature and magnetic field. The temperature dependence of the coercive field $H_C$ and the position of the THE maximum $H_T$ are show as green and red symbols, respectively.*

**Table 1. Comparison of topological Hall effect amplitudes in various materials systems**

| Material system | Maximum THE (μΩ.cm) | Ref. | Note |
|---|---|---|---|
| MnSi | 0.0045 | [23] | Skyrmion system |
| MnP | 0.01 | [26] | Fan spin structure |
| MnGe | 0.16 | [27] | Skyrmion system |
| $SrRuO_3/SrIrO_3$ | 0.2 | [28] | Interface system |
| $Fe_{0.7}Co_{0.3}Si$ | 0.5 | [29] | Skyrmion system |
| EuO | 6 | [30] | Centrosymmetric |
| $Ca_{0.99}Ce_{0.01}MnO_3$ | 120 | This work | Correlated oxide |

The maximum amplitude of the THE is plotted as a function of temperature in Fig. 2e (the corresponding dependence for the AHE is displayed in Fig. 2d), and shows a nearly monotonic decrease as $T$ is raised. Combining the THE data for all temperatures, we build the ($T$, $H$) phase diagram displayed in Fig. 2f that also presents the dependence of the coercive field (extracted from the AHE and the Kerr ellipticity) and the field at which the THE is maximum. The THE is highest in a ($T$, $H$) pocket centred near 1-1.2 T and extending to about 80 K. Beyond this temperature, only the AHE is present, up to the $T_C$ near 100 K.

As initially proposed by Bruno *et al*[8], a THE arises when an electron moves in a medium with spatially varying spin texture, which endows it with a Berry phase. The effect of this Berry phase can be mapped onto that of an effective (emergent) perpendicular magnetic field ***b*** that produces a Hall effect, just as an external magnetic field. The amplitude of ***b*** depends on the topology of the spin textures through the density of the corresponding topological charge. For instance, in a skyrmion array $\langle b \rangle = \Phi_0/d_{sk}^2$ where $\Phi_0$ is the flux quantum and $d_{sk}$ the average distance between skyrmions carrying a topological charge $Q = 1$. In this simple model for strong ferromagnetic exchange coupling, the THE then writes

$$\rho_{THE} = \frac{P\langle b \rangle}{en} = \frac{\Phi_0 P}{d_{sk}^2 en} \qquad (2)$$

with $P$ the spin polarization and $e$ the electron charge.

To gain more insight into the nature of the spin configurations responsible for the THE in our films, we have performed magnetic force microscopy measurements as a function of temperature and out-of-plane magnetic field. Fig. 3a-f show examples of such images taken at negative magnetic field, after saturating with a positive field of +3 T at 10 K. At -0.6 T (Fig. 3a), the magnetization is still fully aligned in the positive field direction and the MFM image shows a homogeneous red contrast. As H increases further towards large negative values, magnetization start to reverse and domains with negative magnetization appear (Fig. 3b). Near -1.2 T the image shows a large density of small bubble-like domains with positive magnetization surrounded by a negative magnetization background (Fig. 3d). Their density progressively decreases as the field increases further to large negative values (cf. Fig. 3e taken at -1.4 T) and they vanish beyond -1.8 T (Fig. 3f). The bubbles visible in Fig. 3c-e have a typical size of ~200 nm, which is comparable to the size of topological bubbles detected in other centrosymmetric manganites[14–16]. By analysing the images collected at different fields, one can extract the field dependence of the number of bubble-like domains ($n_b$). It is plotted in Fig. 3g (right axis) and compared with the field dependence of the THE (left axis). As already revealed by the images, $n_b$ strongly varies with $H$ and, remarkably, peaks near the same field $H_T$ as the THE. The same analysis was done at 40 K (Fig. 3h) and 80 K (Fig. 3i). Similarly, the THE and $n_b$ peak near the same $H_T$ and the value of $H_T$ decreases with temperature for both quantities. This strongly suggests that the THE is created by the specific spin texture associated with the presence of these bubbles.

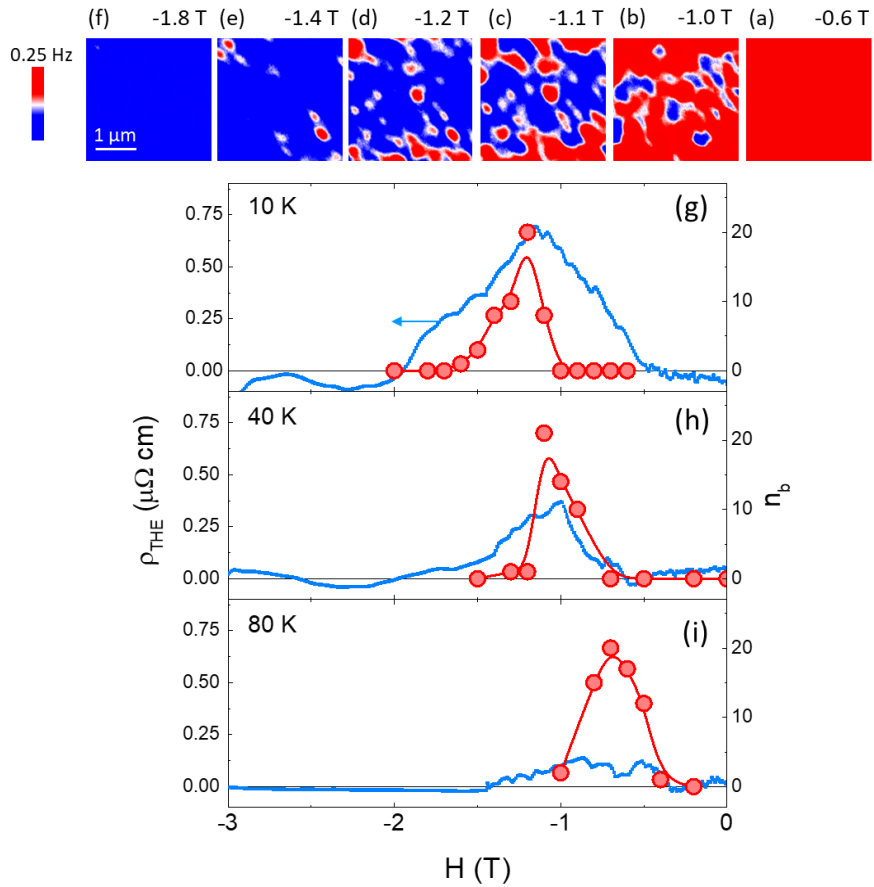

**Figure 3. Connection between micromagnetism and topological Hall effect.** *(a-f) Magnetic force microscopy images at 10 K after applying a positive perpendicular field of +3T, at different negative*

We now discuss in more detail the origin of the AHE and the THE in our films, starting with the AHE. In CMR manganites (CMR: colossal magnetoresistance) such as $La_{0.7}Sr_{0.3}MnO_3$ or $La_{0.7}Ca_{0.3}MnO_3$ the temperature dependence of the AHE is peculiar. It is vanishingly small at low temperatures, developing only when temperature increases to about $T_C/2$ and peaking near $T_C$ (Ref. [31–33]). Two related models[34,35] have been proposed to explain this unusual behaviour. Both consider that the AHE arises from the emergence of dynamic non-coplanar spin configurations having a topological character and describable as skyrmion strings (beginning at a skyrmion and ending at an antiskyrmion). As in the model of Bruno *et al*[8] for the THE, itinerant electrons travelling through these spin textures acquire a Berry phase corresponding to an emergent magnetic field, generating a Hall effect, i.e. $\rho_{AHE} = \frac{P(b)}{en}$ as in Eq. (2). However, the strings are mostly oriented randomly and their contributions practically cancel out. A slightly preferred orientation only arises through spin-orbit coupling. In CMR manganites at low temperature the string density is vanishingly small (magnetization is saturated with all spins collinear) and increases strongly close to $T_C$ (Ref. [35]). As a result, the AHE shows a maximum near $T_C$ [34,35] as in the experiments. In our data (Fig. 2d) this maximum is also visible, but they also evidence a strong AHE at low temperature. Further theoretical work thus appears necessary to accurately describe the AHE in weakly ferromagnetic manganites such as CCMO and in particular its large amplitude below ~20 K, perhaps reflecting a finite density of non-coplanar spin configurations already at low temperature.

Let us now turn to the THE. As the THE occurs in the field range where the micromagnetic corrugation is the largest, one possibility could be that it arises from non-coplanar spin configurations at domain walls, with the same mechanism of the AHE that arises from non-coplanar spin configurations in the domains. Given the relatively high anisotropy of our CCMO films, we estimate that domain walls are rather narrow, with a width $d_{dw}$ in the 10-20 nm range, and thus generate angles between spins on the order of $\theta_{dw} = 180 \times a/d_{dw} = 3\text{-}6°$. This is comparable to the canting angles in the saturated state so that, locally, the AHE should not be much larger at the domain walls than in the domains. In view of their small volume fraction, domain walls should thus not generate an extra Hall effect with an amplitude as large as that of the observed THE by this mechanism.

While the domain walls around the bubbles cannot explain the THE, we argue below that the bubbles themselves can. Magnetic bubbles are characterized by a topological charge $Q$ (i.e. winding number or vorticity) that can be a finite integer or zero depending on the internal structure of the domain wall enclosing the bubble (Ref. [7,13,14]). When the domain wall is a simple Néel or Bloch wall, bubbles have a spin structure very similar to that of skyrmions (the main difference is that bubbles have an extended core, while skyrmions have a point-like core) with the same topological charge of $Q = 1$, Ref. [7]. In other words, such bubbles have a topological character and in the literature they are called "skyrmion-bubbles"[7,36]. If, however, the domain wall comprises one or more pairs of Bloch lines, depending on their structure, $Q$ may be reduced to zero[13]. From the magnetic anisotropy and exchange parameter[17], we find that for the bubble size range observed in Fig. 3, creating a pair of Bloch lines in the domain wall typically increases the bubble energy by an order of magnitude, making their formation unlikely. Thus, most of the observed bubbles in our samples should carry a topological charge of $Q = 1$ just as standard skyrmions and as previously observed in several bulk manganites[14–16]. In this scenario, the skyrmion-bubbles should contribute to the THE through a Berry phase mechanism such as that proposed by Bruno *et al.*[8] However, using Eq. (2) and the average spacing between bubbles detected by MFM, we calculate a THE amplitude weaker by a factor 25-40 than the experimental value. This discrepancy may reflect our inability to detect very small bubbles, although they would then have to be ~25-40 times more abundant that those visible in Fig. 3 ; alternatively, we postulate that the THE in CCMO is amplified by electron correlations, an effect not taken into account in the simple picture of Ref. [8].

To explore this possibility, we have measured the Hall response of CCMO films with different doping levels and carrier densities. As visible in Fig. 4a-d, a THE is also observed at 1%, 2% and 5% and its amplitude varies strongly with Ce concentration. In the Supplementary Material, we show sets of magnetic images for 2% and 5% similar to those of Fig. 3 along with their analysis; just as for the 4% sample, MFM evidences the presence of bubbles whose density with magnetic field dependence tracks that of the THE. Within error bars, the observed bubble size is comparable to that found for 4%.

Fig. 4e summarizes the dependence of the THE amplitude with carrier density. As $n$ decreases, the THE diverges, increasing by three orders of magnitude upon reducing the carrier density by just a factor of ~4. This is in stark contrast with the expected behaviour using Eq. (2).

Rather than a model for the strong coupling regime[8], one applicable to weaker coupling is in fact here better suited since CCMO is a weak ferromagnet. To treat this situation, we consider free electrons exchange-coupled to the static magnetization texture (smoothly varying in space) and subjected to (normal, non-magnetic) impurity scattering, with both adiabatic and non-adiabatic contributions (cf. Supplementary Material and Ref. [37]). In this case, the Hall resistivity is expressed as

$$\rho_{THE} \sim \frac{\langle b \rangle m^* J_{fm}}{n^{5/3}} \qquad (4)$$

where $J_{fm}$ is the exchange coupling between conduction electrons and magnetization multiplied by the ferromagnetic moment, which, as the magnetization, is proportional to the doping level (i.e. to $n$), cf. Fig. 1g (inset). Unlike in the strong coupling model the THE amplitude now depends on the electron effective mass. In a doped correlated insulator such as CCMO, as the transition to the insulating state is approached from the metallic side (here upon reducing the Ce doping level), the effective mass diverges[38], which is the signature of strong correlations[39]. This is documented for several perovskite systems, for instance Sr-doped LaTiO₃ [40] where upon approaching the Mott insulating state of LaTiO₃, $m^*$ increases up to m*/m₀ ≈ 30, or Ce-doped SrMnO₃ (very similar to CCMO) in which m*/m₀ ≈ 10 at 1-2% Ce while the bare mass is 0.6$m_0$ [41].

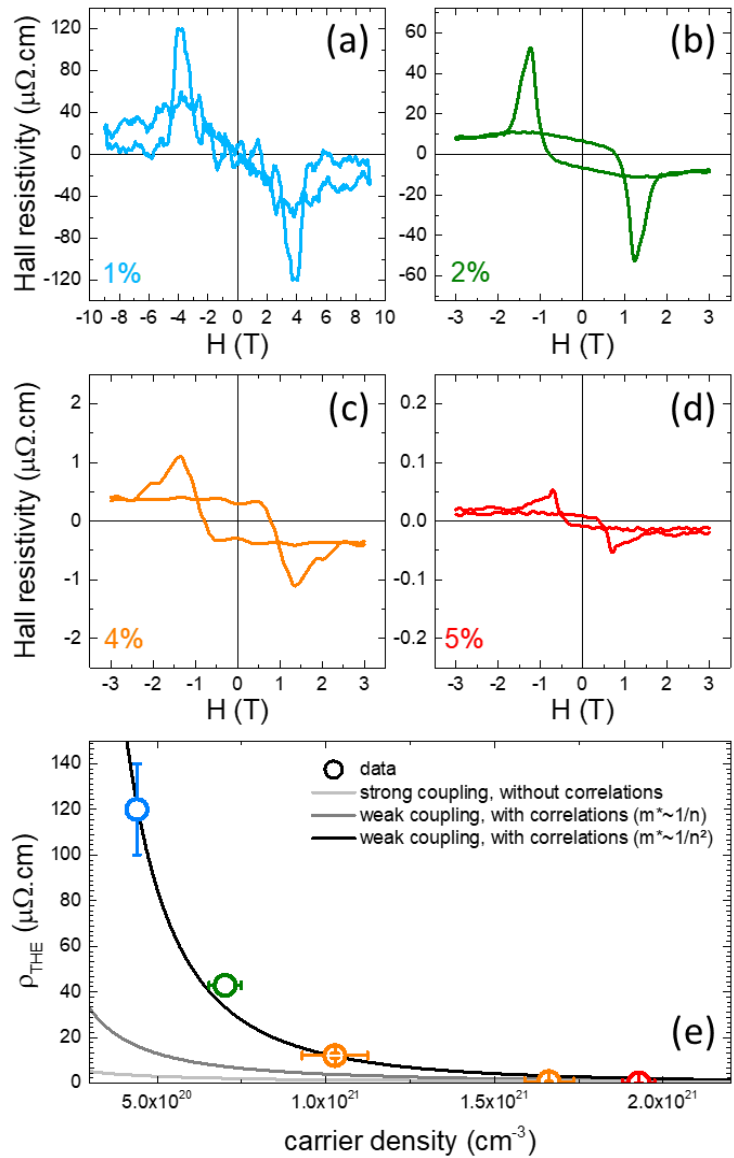

**Figure 4. Doping dependence of the topological Hall effect.** *Topological Hall effect at 20 K for 1% (a) and 15 K for 2% (b), 4% (c) and 5% doping (d). Relationship of the THE (e) with the carrier density.*

Coming back to Eq. (4), we can incorporate the effect of electron correlations by assuming an enhancement of the effective mass. With a scaling of $m^*$ as $1/n^\alpha$ ($\alpha \geq 1$) we see that $\rho_{THE}$ has to scale as $1/n^{2/3+\alpha}$, which provides ground for the strong enhancement of the topological Hall effect at low doping, due to the renormalization of the effective mass by electronic correlations. In Fig. 4e we compare the experimental dependence of the topological Hall resistivity with the carrier density with scaling laws in the strong coupling regime[8] ($1/n$) and in the weaker coupling regime considered here, in the presence of correlations for different values of $\alpha$. While the $1/n$ dependence of the strong coupling model clearly fails to reproduce the data, the agreement is good within the weaker coupling and strong correlations regime, using $\alpha=2$. These results indicate that electronic correlations can highly amplify the topological Hall response and call for a detailed theoretical treatment of this remarkable phenomenon.

In summary, we have observed a very large topological Hall effect in epitaxial thin films of a doped charge-transfer insulating perovskite, $CaMnO_3$. The emergence of the THE is consistently associated with the appearance of magnetic bubbles imaged by magnetic force microscopy, at different temperatures and for several doping levels. The data therefore indicate that these bubbles are the cause of the THE and strongly suggest that, just as skyrmions, these bubbles carry a topological charge. The THE amplitude diverges as carrier density decreases and the correlated insulating state of $CaMnO_3$ is approached, which can be accounted for by considering the enhancement of the electron effective mass by strong correlation effects, within a weak exchange coupling regime. Our results suggest that between conventional ferromagnetic and antiferromagnetic skyrmion systems, weak ferromagnets emerge as promising materials to harbour skyrmions easier to detect by electrical means. Our work also indicate that magnetic perovskite oxides represent an exciting class of materials exhibiting novel transport phenomena stemming from non-trivial spin textures, highly tuneable through doping, strain and interface engineering[28]. In particular, the strong sensitivity of the THE to the carrier density offers interesting perspective for its non-volatile control by electric-field. In heterostructures combining $CaMnO_3$ with ferroelectrics such as $BiFeO_3$ large, non-volatile modulation of the linear Hall effect and longitudinal resistivity have indeed been reported (Ref. [42]).

**Acknowledgements**

We thank V. Cros, V. Dobrosavljevic, J. Iñiguez, J.-V. Kim, D. Maccariello, J. Matsuno, I. Mertig, N. Nagaosa and N. Reyren for useful discussions, J.-Y. Chauleau and M. Viret for second harmonic

generation experiments, J. Varignon for preparing Fig. 1a and J.-M. George for his help with some magnetotransport measurements. This research received financial support from the ERC Consolidator grant "MINT" (contract number n°615759) and the ANR project "FERROMON". This work was also supported by a public grant overseen by the ANR as part of the "Investissement d'Avenir" program (LABEX NanoSaclay, ref. ANR-10-LABX-0035) through projects "FERROMOTT" and "AXION" and by the Spanish Government through Projects No. MAT2014-56063-C2-1-R, and Severo Ochoa SEV-2015-0496 and the Generalitat de Catalunya (2014SGR 734 project). B. C. acknowledges Grant No. FPI BES-2012-059023, R. C. acknowledges support from CNPq-Brazil and J.S. thanks the University Paris-Saclay (D'Alembert program) and CNRS for financing his stay at CNRS/Thales. Work at Rutgers is supported by the Office of Basic Energy Sciences, Division of Materials Sciences and Engineering, U.S. Department of Energy under Award No. DE-SC0008147. HK is supported by JSPS KAKENHI Grant Numbers 25400339, 15H05702 and 17H02929. KN is supported by Grant-in-Aid for JSPS Research Fellow Grant number 16J05516, and by a Program for Leading Graduate Schools "Integrative Graduate Education and Research in Green Natural Sciences".

**Data availability statement**

The data that support the plots within this paper and other findings of this study are available from the corresponding author upon reasonable request.

**Methods**

**Fabrication and structural characterisation.** The $Ca_{1-x}Ce_xMnO_3$ ($x$ = 0, 1, 2, 4, 5% nominal Ce concentrations) 20 nm thin films were grown by pulsed laser deposition from stoichiometric targets on (001) $YAlO_3$ substrates using a Nd:YAG laser. Commercial $YAlO_3$ (001) oriented substrates were prepared with acetone cleaning and ultrasound in propanol, and then annealed at 1000°C in high $O_2$ pressure. The substrate temperature ($T_{sub}$) and oxygen pressure ($P_{O2}$) during the deposition were 620°C and 20 Pa, respectively. Post-deposition annealing was performed at $T_{sub} \approx 580$°C and $P_{O2}$ = 30 kPa for 30 minutes, followed by a cool-down at the same oxygen pressure. 2θ-ω X-ray diffraction scans were performed with a Panalytical Empyrean equipped with a hybrid monochromator for Cu $K_{\alpha1}$ radiation and a PIXcel3D detector. The thickness of the $Ca_{1-x}Ce_xMnO_3$ thin film was measured by X-ray reflectivity with a Bruker D8 DISCOVER. Hall bars for magnetotransport measurements were patterned by optical lithography and argon ion etching. Electrical contacts for measurements were made on platinum electrodes defined by a combination of lithography and lift-off techniques.

**Electrical characterization.** The magnetotransport characterization of the samples was performed in a Quantum Design Physical Properties Measurement System (PPMS) Dynacool. The temperature dependence of the resistivity was measured at a constant current of 5 µA during a warming run after field cooling in a cryostat through a closed liquid helium loop. For Hall measurements, magnetic fields were swept up to ±9 T. To separate the Hall contribution from that of the longitudinal magnetoresistance, an antisymmetrization procedure was performed by separating the positive and negative field sweep branches, and interpolating the two onto the same field coordinates and then antisymmetrizing using

$$\rho'_+(H) = [\rho_+(H) - \rho_-(-H)]/2$$

$$\rho'_-(H) = [\rho_-(H) - \rho_+(-H)]/2.$$

**Magnetic characterization.** Superconducting Quantum Interference Device (SQUID) magnetic characterization of the samples was performed in a Quantum Design Magnetic Properties Measurement System (MPMS). The temperature dependence of the magnetization was measured at a constant field of 1 kOe during a warming run after field cooling in a cryostat with liquid helium. The magnetization loops were obtained by sweeping the magnetic field up to ±5.5 T.

**Magnetic force microscopy.** The MFM experiments were carried out in a homemade cryogenic atomic force microscope (AFM) using commercial piezoresistive cantilevers (spring constant $k \approx 3$ N/m, resonant frequency $f_0 \approx 42$ kHz). The homemade AFM is interfaced with a Nanonis SPM Controller (SPECS) and a commercial phase-lock loop (OC4)[43]. MFM tips were prepared by depositing nominally 100 nm Co film onto bare tips using e-beam evaporation. MFM images were taken in a constant height mode with the scanning plane ~60 nm above the sample surface. MFM signal: the change of cantilever resonant frequency, is proportional to out-of-plane stray field gradient[44]. Electrostatic interaction was minimized by nulling the tip-surface contact potential difference. Red (blue) regions in MFM images represent up (down) ferromagnetic domains, where magnetizations are parallel (anti-parallel) with the positive external field. Using the ImageJ software, all images were binarised at the same threshold, starting from the same vertical scale. Then we used the particle analysis module of this software to count the number of bubbles and estimate their size.

**Magneto-Optical Kerr measurements**. Magneto-optical experiments were done using light from a 150 W Xe arc lamp, which was dispersed by a monochromator, collimated, and then linearly polarized, by the action of a Glan-Thompson prism, which was rotated by 45° with respect to the modulator axis of a photoelastic modulator (PEM). After reflection on the sample surface, the light goes towards an analyzer that can be set at two different angles with respect to the PEM axis, namely, 0° and 90°, to record the magneto-optical signals of s- and p-polarized light, respectively. In our experiments, Kerr ellipticity (ε) was measured in polar configuration with s- and p-polarized light incident at angles close to the normal to the surface. The signal is collected from the detector and brought to a lock-in amplifier synchronized to the frequency $\Omega$ of the PEM retardation angle. In this optical arrangement, the ellipticity is given by $\varepsilon = \frac{1}{4cJ_1(\varphi_0)}\frac{I_\Omega}{I_0}$, where $I_\Omega$ is the first harmonic of the intensity recorded at the detector and $I_0$ is the background intensity measured with a dc-multimeter. The calibration constant $c$ is determined experimentally.

## Author contributions